\documentclass[journal=nanolett,manuscript=article]{achemso}

\usepackage[version=3]{mhchem} 
\usepackage{graphicx}
\usepackage{dcolumn}
\usepackage{bm}
\usepackage[latin1]{inputenc}
\usepackage{textcomp}
\usepackage{amsmath}
\usepackage{color}
\usepackage{ulem}
\usepackage[final]{pdfpages}

\author{Wei Fang}
\affiliation{Thomas Young Centre, London Centre for Nanotechnology, and Department of Physics and Astronomy, University College London, London WC1E 6BT, UK}
\alsoaffiliation{Laboratory of Physical Chemistry, ETH Zurich, CH-8093 Zurich, Switzerland}
\author{Kastur M.\ Meyer auf der Heide}
\affiliation{Ruhr-Universit\"{a}t Bochum, Lehrstuhl f\"{u}r physikalische Chemie\,I, Universit\"{a}tsstr.\,150, D-44801~Bochum, Germany}
\author{Christopher Zaum}
\affiliation{Leibniz Universit\"at Hannover, Institut f\"ur Festk\"orperphysik, Appelstr.\ 2, D-30167 Hannover, Germany}
\author{Angelos Michaelides}
\email{am452@cam.ac.uk}
\affiliation{Yusuf Hamied Department of Chemistry, University of Cambridge, Lensfield Road, Cambridge CB2 1EW, UK}
\alsoaffiliation{Thomas Young Centre, London Centre for Nanotechnology, and Department of Physics and Astronomy, University College London, London WC1E 6BT, UK}
\author{Karina Morgenstern}
\email{karina.morgenstern@rub.de}
\affiliation{Ruhr-Universit\"{a}t Bochum, Lehrstuhl f\"{u}r physikalische Chemie\,I, Universit\"{a}tsstr.\,150, D-44801~Bochum, Germany}

\title{Rapid water diffusion at cryogenic temperatures through an inchworm-like mechanism}


\begin{document}

\begin{tocentry}




\includegraphics[width=8.2cm]{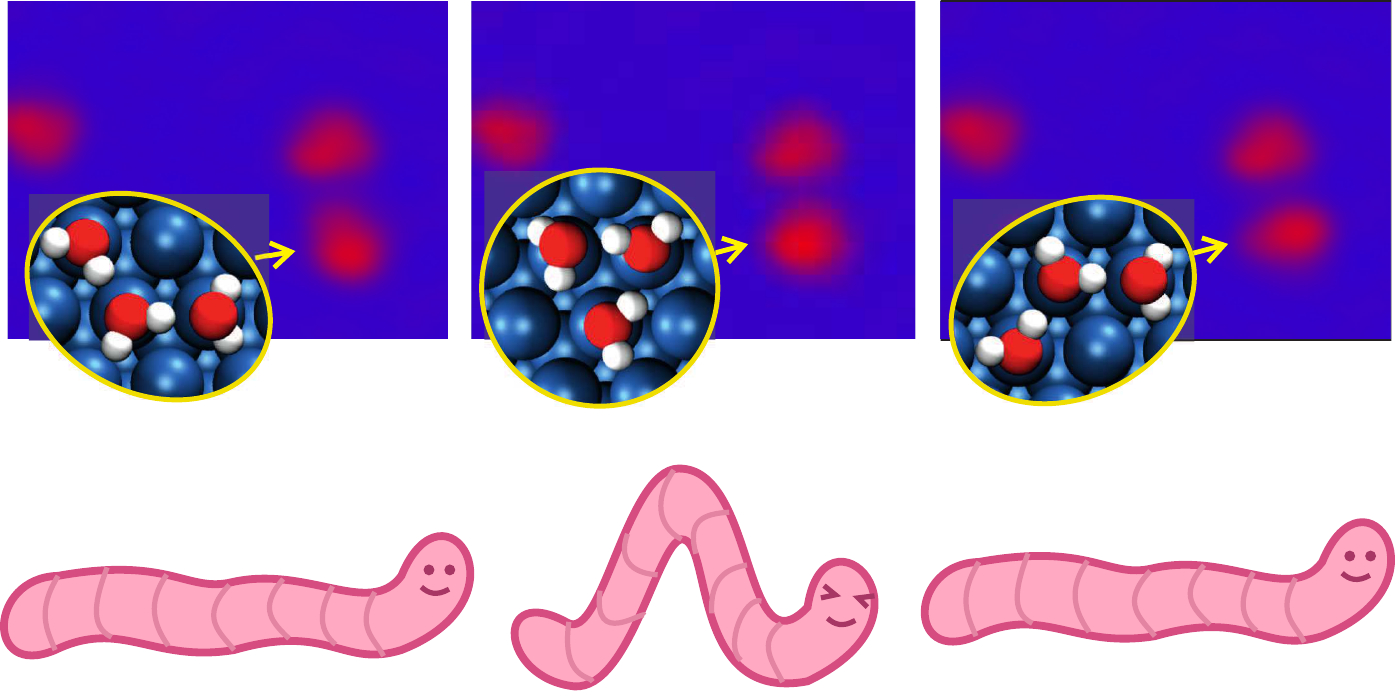}

\end{tocentry}


\begin{abstract}
Water diffusion across the surfaces of materials is of importance to disparate processes such as water purification, ice formation, and more. 
Despite reports of rapid water diffusion on surfaces the molecular-level details of such processes remain unclear. 
Here, with scanning tunneling microscopy, we observe structural rearrangements and diffusion of water trimers at unexpectedly low temperatures ($<$10 K) on a copper surface; temperatures at which water monomers or other clusters do not diffuse. 
Density functional theory calculations reveal a facile trimer diffusion process involving transformations between elongated and almost cyclic conformers in an inchworm-like manner. 
These subtle intermolecular reorientations maintain an optimal balance of hydrogen-bonding and water-surface interactions throughout the process. 
This work shows that the diffusion of hydrogen-bonded clusters can occur at exceedingly low temperatures without the need for hydrogen bond breakage or exchange; findings that will influence Ostwald ripening of ice nanoclusters and hydrogen bonded clusters in general.
\end{abstract}

\maketitle


\section{Introduction}

The operation of desalination membranes or magnesium-air batteries, the formation of clouds or the fracture of rocks by ice, skiing, skating or the impact of raindrops on glass -- these and countless other everyday technological and scientific processes involve the motion of water molecules across the surfaces of inorganic materials. 
This broad importance has motivated a large body of work aimed at the understanding  of interfacial water. 
At the fundamental molecular-level much of this work has focused on the structure and dynamics of interfacial and confined water \cite{HENDERSON2002,carrasco12,chemrev.6b00045,maier15,siria_new_2017}. %
However, despite this large body of work -- from which we have learned a great deal 
about the fundamental nature of water at interfaces -- understanding of how water moves
across surfaces remains incomplete.
This applies to both the flow of liquid water and the motion of individual water molecules or clusters.
Whether it is water flowing through a nanoscale channel or individual water clusters diffusing across a surface, the motion depends on a balance of water-water and water-surface interactions. 
However, understanding these interactions in detail remains a challenge.

However, by examining well-defined atomically precise surfaces 
at cryogenic temperatures it is possible to track the motion of individual water molecules and water clusters.
In particular, with scanning tunneling microscopy (STM), detailed and fascinating 
results for water motion on the surfaces of metals and oxides have been obtained (see e.g. refs. \cite{mitsui02,maier15,Marx_ZnO,NaCl_vdW}).
Notably, in a seminal contribution from Salmeron and co-workers \cite{mitsui02},
it was shown that on a Pd surface water dimers diffuse more rapidly than water monomers.
This result was surprising because of the larger mass of the dimer and also the stronger interaction strength of the dimer with the surface; often diffusion barriers of adsorbates on surfaces are proportional to the strength
of the adsorption bond \cite{nilekar06,SEEBAUER1995265,YildirimJPCC2012}.

These observations have been explained with the help of density functional theory (DFT) and a mechanism for rapid dimer motion involving a tunneling assisted hydrogen bond exchange process \cite{ranea04,dimersurf}. 
Subsequent studies on other metal surfaces showed that on Cu(111) monomers and dimers have similar diffusion rates, whereas on Ag(111), monomers are the faster species \cite{Cu_exp,heidorn18}. 
What happens beyond the dimer remains largely unexplored,
although clusters larger than the dimer also diffused rapidly in the Pd experiments \cite{mitsui02}, and water clusters of unknown size have been shown to be involved in the formation of ice \cite{heidorn16,heidorn18}.

In this article we go down to 5 K where almost all adsorbate motion is frozen out to explore in detail the motion of water clusters on a surface. 
%
Through a combination of painstaking time-lapsed STM and DFT we find that of the species on the Cu(111) surface studied, water trimers have the highest mobility: changing conformation and diffusing at temperatures as low as ca.\ 6 K.
Notably, frequent changes between elongated and compact trimer structures promote, together with facile rotation of the elongated form, rapid trimer diffusion; diffusion which exceeds that of water monomers and all other adsorbed water clusters observed on the surface.
In contrast to the case of rapid dimer motion on Pd, the rapid motion of the trimer is achieved through delicate changes of internal structure and does not require the (tunneling assisted) donor acceptor (DA) exchange of hydrogen bonds. 
%


 \section{Materials and Methods}
Our STM measurements were performed under ultrahigh vacuum (UHV) conditions with a variable low-temperature STM, calibrated to sub-Kelvin precision between 5~K and 50~K \cite{zaum16}. 
The Cu(111) surface was cleaned by repeated cycles of sputtering (Ne$^{+}$, 4 to $5 \cdot 10^{-5}$~mbar, 1 to 2~$\mu$A, 1.3~keV, 40~min) and annealing (873~K, 8~min). 
Water (D$_2$O) of milli-Q quality is filled into a glass tube that is connected via a leak valve to a vacuum chamber that is separated from the preparation chamber by a gate-through valve.
After purification through several freeze-pump-thaw cycles, a pressure of $1.5\cdot 10^{-7}$~mbar is established in the molecule deposition chamber via a leak valve from the room temperature water vapor in the glass tube. 
The sample is either exposed for 45~s at 15 K or for 6.4~s at 81~K.
In the former case, the sample resides within the cold shields surrounding the STM, reducing the deposition rate to around $ 10^{-3}$ BL/min leading to a coverage around $6\cdot 10^{-4}$ BL (bilayer, corresponding to a buckled monolayer at a density of two water molecules per three Cu surface atoms). 
In the latter case, it is on a nitrogen-cooled manipulator close to the water inlet, where the effective pressure at the sample is around $10^{-9}$~mbar, yielding a coverage of around $5\cdot 10^{-3}$ BL. 
At 15~K monomers are immobile, while at 81~K, these and other oligomers are mobile on Cu(111) \cite{bertram19}, but rearrangements of water molecules within amorphous water structures is inhibited \cite{mehlhorn07}.
Dynamical changes are investigated in time-lapsed series between 5~K and 12 K as detailed in \cite{morgenstern04_Cudimer}.

The density-functional theory (DFT) calculations were carried out using the Vienna Ab initio Simulation Package (VASP) \cite{kresse_efficient_1996} with the optB88-vdW functional \cite{optB88}. 
A plane-wave cutoff of 500~eV was used.
The metal surfaces were represented using a slab with 4 layers of Cu in a 5$\times$4 unit cell with a 3$\times$3$\times$1 K-point mesh. 
A vacuum of at least 1.4 nm was placed above each surface. 
The climbing image nudged elastic band (CI-NEB) \cite{CINEB} method was used to obtain the potential energy barriers and minimum energy pathways. 
The force convergence criteria for the geometry optimizations and CI-NEB calculations was 0.01 eV/\AA.
Since surface relaxation effects during the diffusion processes are small, the substrate atoms were fixed at the relaxed clean surface geometry during optimizations while the water molecules were fully relaxed.
These computational settings are consistent with those used in our recent studies on related systems \cite{bertram19,dimersurf}. 
We present in the supporting information (SI) sections S.VII and S.VIII for tests on the sensitivity of the results to various choices in computational settings, as well as an applied electric field. 

 \section{Results and discussion}
\begin{figure}[htb]
\includegraphics[width=0.99\columnwidth]{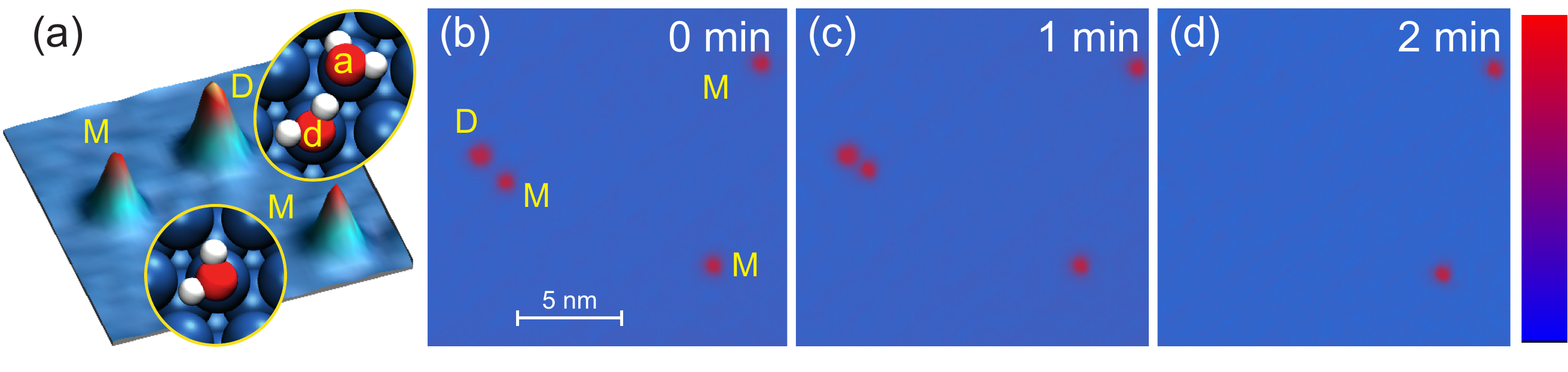}
\caption{Water monomer and dimer adsorption and trimer formation. (a) 3D view of an STM image showing two monomers (M) and a  dimer (D) along with the calculated structures; $d$ and $a$ mark the water molecules in the dimer acting as hydrogen bond donor and acceptor, respectively. 
(b to d) Time sequence of STM images showing a monomer and dimer approaching (panels b to c) and subsequently disappearing from view once a mobile trimer forms (panel d).
A false color-scale from blue to red is used and the following imaging conditions were employed: 24~K, 25~mV, 20~pA.
\label{image_trimform}} 
\end{figure}

We begin by demonstrating the high mobility of water trimers on Cu(111) compared to water monomers and dimers.
Fig.\ \ref{image_trimform}a displays an STM image of water monomers (M) and a water dimer (D) on Cu(111). 
As shown before, \cite{bertram19} both species image as bright circular protrusions, with 
the dimer having a slightly larger diameter. 
The adsorption geometry of each species has been characterized before \cite{Michaelides01,ranea04,Meng2004,Carrasco_2009_1,carrasco12,C0CP00994F} and is shown as insets in Fig.\ \ref{image_trimform}a. 
Monomers adsorb at atop sites of Cu(111). 
In the dimer, one of the molecules (the hydrogen bond donor, $d$) adsorbs in a similar atop site configuration, while the other molecule (the hydrogen bond acceptor, $a$) interacts weakly with the substrate and rotates easily around the surface-bound molecule. 
It is because of this facile rotation that the dimer also images as a circular protrusion, but with a larger diameter than the monomer. 
%

The diffusion barriers of the monomer and dimer were previously determined to be 75 ($\pm$ 4) meV and 80 ($\pm$ 8) meV, respectively, from time-lapsed STM diffusivity studies \cite{bertram19}.
From DFT calculations these barriers were predicted to be 86 and 89 meV (including harmonic zero point energy (ZPE) corrections) for the monomer and dimer, respectively, and thus close to the experimental values. 
DFT also showed that the most favorable process for both monomer 
and dimer diffusion was a simple hopping process from atop to atop sites (see SI section S.X).
At 24 K and on the timescale of the measurements shown in Fig.\ \ref{image_trimform} both species diffuse across Cu(111). 
In the sequence of images shown in panels (b) to (d) a monomer and dimer approach each other and merge to form a new species; a trimer which immediately disappears from view and is absent from the region of the surface imaged in panel (e) (see supplementary movie). 
Thus, trimer diffusion is considerably more facile than monomer and dimer diffusion, implying a lower diffusion barrier for the trimer than the two other species.

\begin{figure}[htb]
\includegraphics[width=0.99\columnwidth]{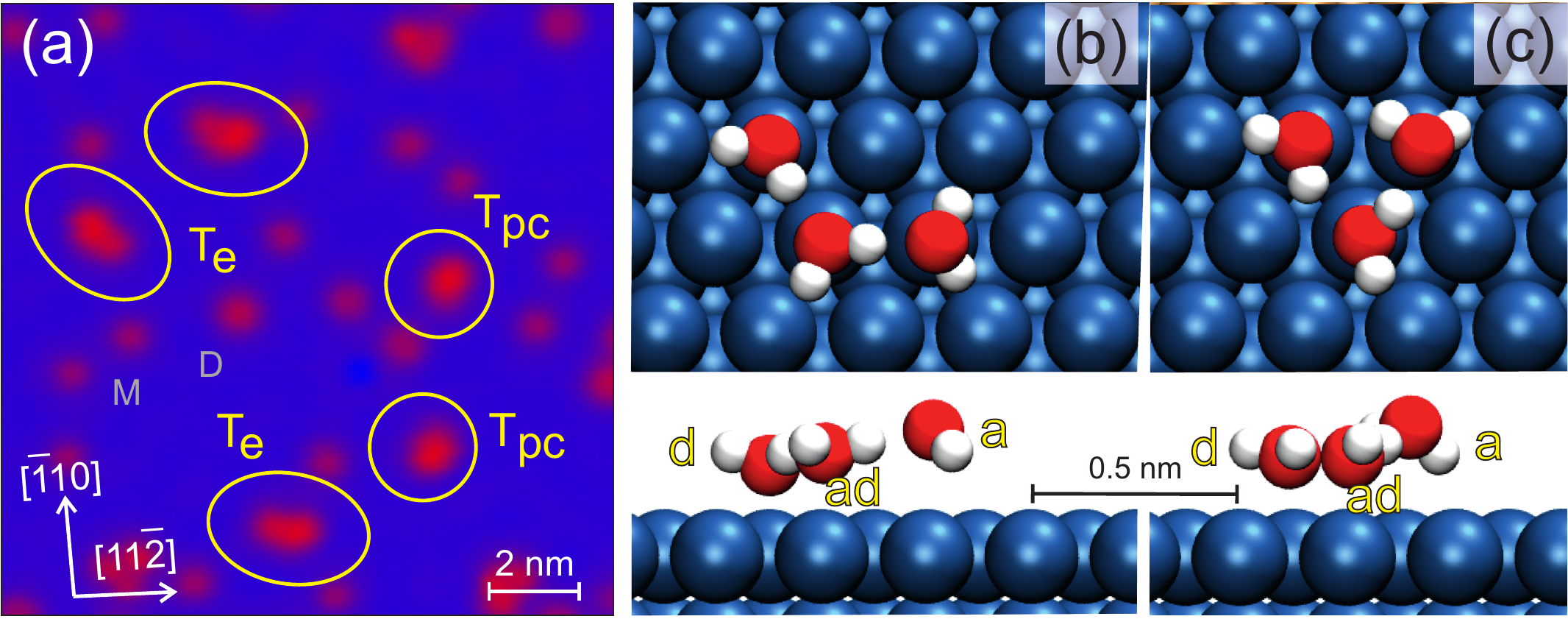}
\caption{Water oligomers on Cu(111). 
(a) STM image  with various species indicated: 
Pseudo-cyclic trimers marked by circles (T$_\text{pc}$); Elongated trimers marked by ellipses (T$_\text{e}$); for reference a monomer (M) and; a dimer (D) are also shown.
The imaging conditions were: 5.2~K, 65~mV, 24~pA.
(b - c) Lowest-energy configurations of the elongated trimer (T$_\text{e}$, panel b) and the pseudo-cyclic trimer (T$_\text{pc}$, panel c) shown from top and side views. 
The labels $a$ and $d$ mark water molecules acting as hydrogen-bond acceptors and donors, respectively. 
\label{image_overview}} 
\end{figure}

In order to deepen understanding of the trimer and the diffusion process we carried out STM measurements at different coverages and temperatures and used DFT to aid the interpretation. 
Fig.\ \ref{image_overview} shows the surface at a lower temperature of 5 K and at a slightly higher water coverage, where clusters of up to approx.\ 10 molecules are observed.
As noted earlier, the smallest round protrusions are monomers and the next largest protrusions dimers.
Monomers and dimers are centro-symmetric, in contrast to the next largest protrusions which are the focus of the current study.  
These slightly larger clusters are either circular (referred to as pseudo-cyclic, PC) or elongated (E).
The circular protrusions are, at a half-width of 1 nm, around ten percent broader than the round protrusions of the dimers.
In addition, each exhibits a small off-center protrusion of greater apparent height.
Similarly, also each elongated protrusion exhibits a small protrusion of greater apparent height (Fig.\ \ref{image_overview}a,b).
The long axes of the elongated protrusions roughly follow the $\langle 112\rangle$ directions of Cu(111) (Fig.\ \ref{image_overview}a).


To learn more about the structures of the trimers we did a set of DFT calculations for adsorbed water trimers, taking into consideration adsorbed trimers with different hydrogen-bonding arrangements. 
From this we concluded that the elongated structure is a trimer with a wide internal angle (Fig.\ \ref{image_overview}b), whereas the circular trimer is a pseudo-cyclic structure (Fig.\ \ref{image_overview}c). 
The elongated trimer structure shown in Fig.\ \ref{image_overview}b
is the lowest energy adsorbed trimer structure identified. 
(See the SI section S.III for other metastable trimer structures, including a structure only 10 meV less stable with a different chirality.) 
The adsorption energy of the trimer is 1.0 eV, relative to a gas phase trimer.
This is larger than the equivalent adsorption energies of the water dimer (0.68 eV) and water monomer (0.32 eV).
As with the monomer and dimer, water molecules are located at atop sites with one molecule acting as a hydrogen bond donor ($d$), one as a hydrogen bond acceptor ($a$), and the central molecule acting as both an acceptor and donor ($ad$). 
The pseudo-cyclic structure is only 31 meV less stable than the lowest energy elongated structure. 
In this structure the water molecules are again at atop sites and we refer to it as pseudo-cyclic because the water-water distances within the trimer ring are not equal and one water sits higher above the surface than the other two.
Indeed, using a standard definition of hydrogen-bonding \cite{Luzar_Nature_1996}, we conclude 
that there are just two hydrogen-bonds in the PC structure and we label the molecules accordingly as {$d$, $a$, and $ad$ in Fig.\ \ref{image_overview}b}. 
We discuss in the SI section S.VI that this finding is not very sensitive to the choice of H-bond definition.
Note that the pseudo-cyclic structure is 51~meV 
more stable than its cyclic counterpart despite having one H-bond less, because there is less geometric deformation (strain) in the case of the pseudo-cyclic cluster. 
In both the elongated and pseudo-cyclic trimers the $a$ molecule has its hydrogen bonds pointing towards the surface and sits higher above the surface than the other molecules, giving rise to the protrusion of larger apparent height in the STM images.
The two lowest energy trimers thus reflect the symmetry of the two experimentally observed structures.
%


\begin{figure}[htb]
\includegraphics[width=0.9\columnwidth]{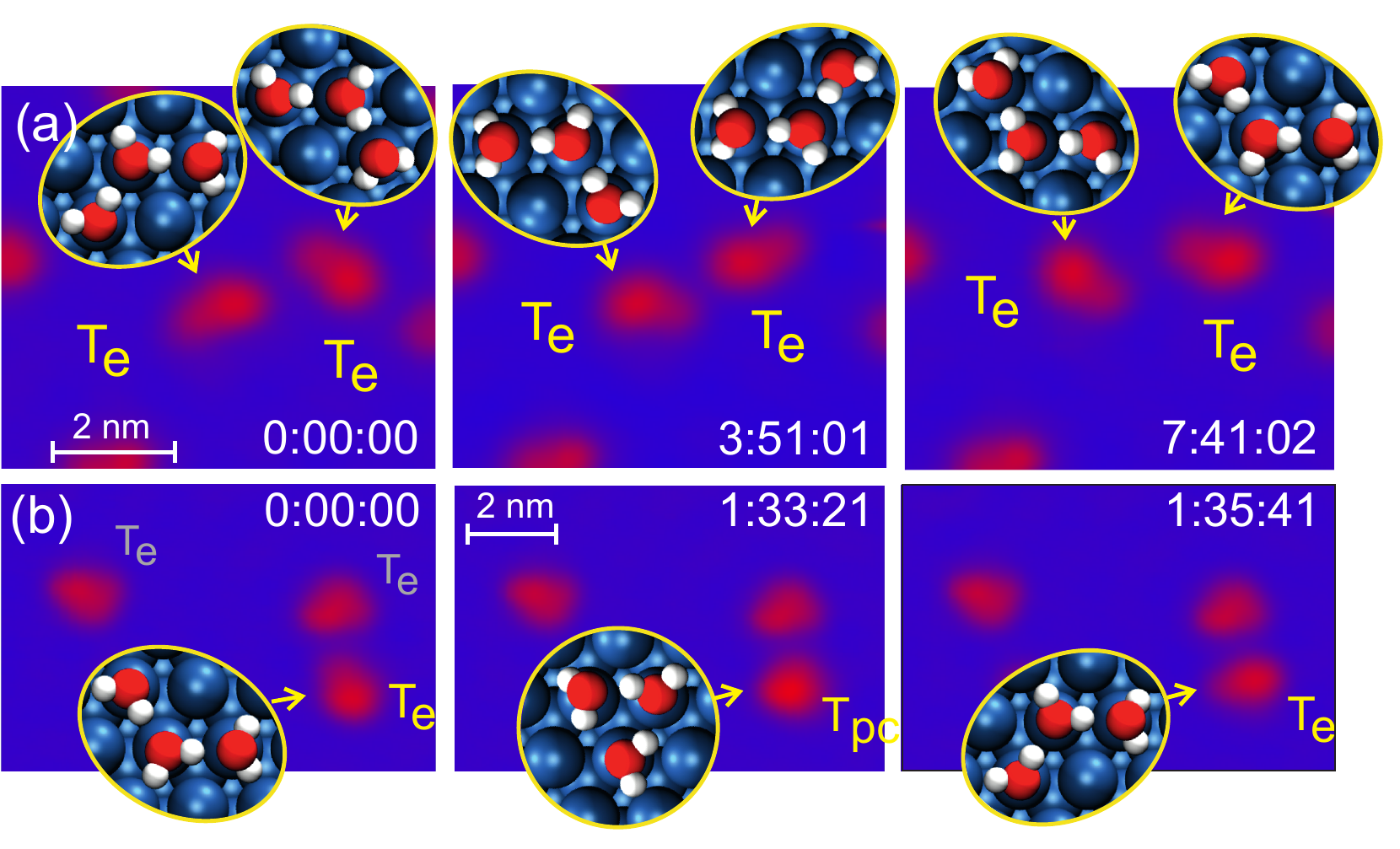}
\caption{
Rotation and interconversion between water trimers at 5-7 K. 
(a) Time series showing two elongated trimers (T$_\text{e}$) reorientating between different rotational configurations. 
(b) Time series showing the conversion from a T$_\text{e}$ trimer to a T$_\text{pc}$ (pseudo-cyclic) trimer and back again to a  T$_\text{e}$ trimer. 
The series in (a) was recorded at 7~K, with the following tunneling conditions: 65~mV, 24~pA.  
The series in (b) was recorded at 5.1~K (first image) to 5.8~K (two later images), with the following tunneling conditions: 65~mV, 24~pA. The times indicated in (a) and (b) are in h:min:s.
\label{image_changeorientation}}
\end{figure}

To probe the dynamics of the trimer we varied the temperature in the 5 to 10 K temperature regime and imaged the surface for extended time periods. 
Interestingly we find that upon very slightly increasing the temperature (by less than 1.0 degree) motion of the water trimers is observed. 
Specifically, the elongated trimer interconverts between different configurations. 
This is seen by imaging a pair of adsorbed trimers for almost 8 hours at 7~K (Fig. \ref{image_changeorientation}a and SI for a movie).
After this time, the trimer on the left is in a different orientation at the same chirality, while the trimer on the right is in the same orientation but with different chirality, with the brighter protrusion changing its position upwards on the image with respect to the less bright one.
The surface symmetry of an fcc(111) surface facilitates six rotamers of each of the two enantiomers for the elongated trimer; there should thus exist twelve different forms.
As shown in the SI section S.II all twelve forms have indeed been imaged. 
In addition to the rotation process we also see interconversion between the elongated and pseudo-cyclic trimer structures.
The elongated trimers in Fig. \ref{image_changeorientation}b are static for more than an hour at 5.1 K.
Upon raising the temperature to 5.8 K, the one at the lower left converts into a pseudo-cyclic trimer 
after approximately 90 minutes and back to an elongated trimer approximately 2 minutes later.
Thus, an interconversion between the elongated and pseudo-cyclic trimer is possible starting at temperatures as low as around 6~K. 
The greater abundance of elongated trimers and the short lifetime of the pseudo-cyclic trimers are in line with the calculated higher stability of the elongated trimer.

Having identified the most stable trimers and seen that the trimers are mobile and can interconvert we now use DFT to examine mechanisms for diffusion and exchange between the different conformers. 
We began by computing the barrier for the most stable trimer to translate across the surface through a straightforward hopping mechanism.
The lowest energy pathway for this process has a barrier of 118~meV (see the SI section S.V for details of the mechanism).
This barrier is larger than the barriers for the monomer and dimer, which as noted earlier are approximately 75 - 90 meV, from both experiment and DFT calculations \cite{bertram19}.
Given that water dimer motion through a H-bond exchange mechanism is possible on surfaces \cite{mitsui02,ranea04,dimersurf}, we explored various trimer diffusion processes in which H-bonds are broken and reformed.
These calculations revealed that for the trimer on Cu(111) all H-bond exchange processes considered had relatively high barriers of $\geq$300 meV (SI section S.IV). 
Thus H-bond exchange processes do not explain the motion observed. 

\begin{table}[!ht]
    \centering
    \begin{tabular}{l|cccc}
    \hline
    \hline
         & Inchworm & ~~T$_\text{e}$ translation~~ & ~~T$_\text{e}$ rotation~~ & ~~T$_\text{e}$ to T$_\text{pc}$~~ \\ \hline
        ``Standard'' Barrier (meV) & 82 & 118 & 35 & 31 \\ 
        ``Corrected'' Barrier (meV) & 75 & 108 & 33 & 43 \\ 
        \hline
        \hline
    \end{tabular}
    \caption{
    DFT computed activation energy barriers for the key processes discussed in the text. Specifically, the following processes are considered: The overall inchworm diffusion process shown in Fig. 4; A simple trimer translation process; Rotation of the T$_\text{e}$ trimer; Transition from the T$_\text{e}$ trimer to the T$_\text{pc}$ trimer. Barriers are reported using the ``standard'' computational set-up (as described in the methods section) and ``corrected'' to take into account zero point energy, surface relaxation, and finite size effects, as detailed in SI section S.VII. 
    }
    \label{barrier}
\end{table}

\begin{figure}[!ht]
\includegraphics[width=0.98\columnwidth]{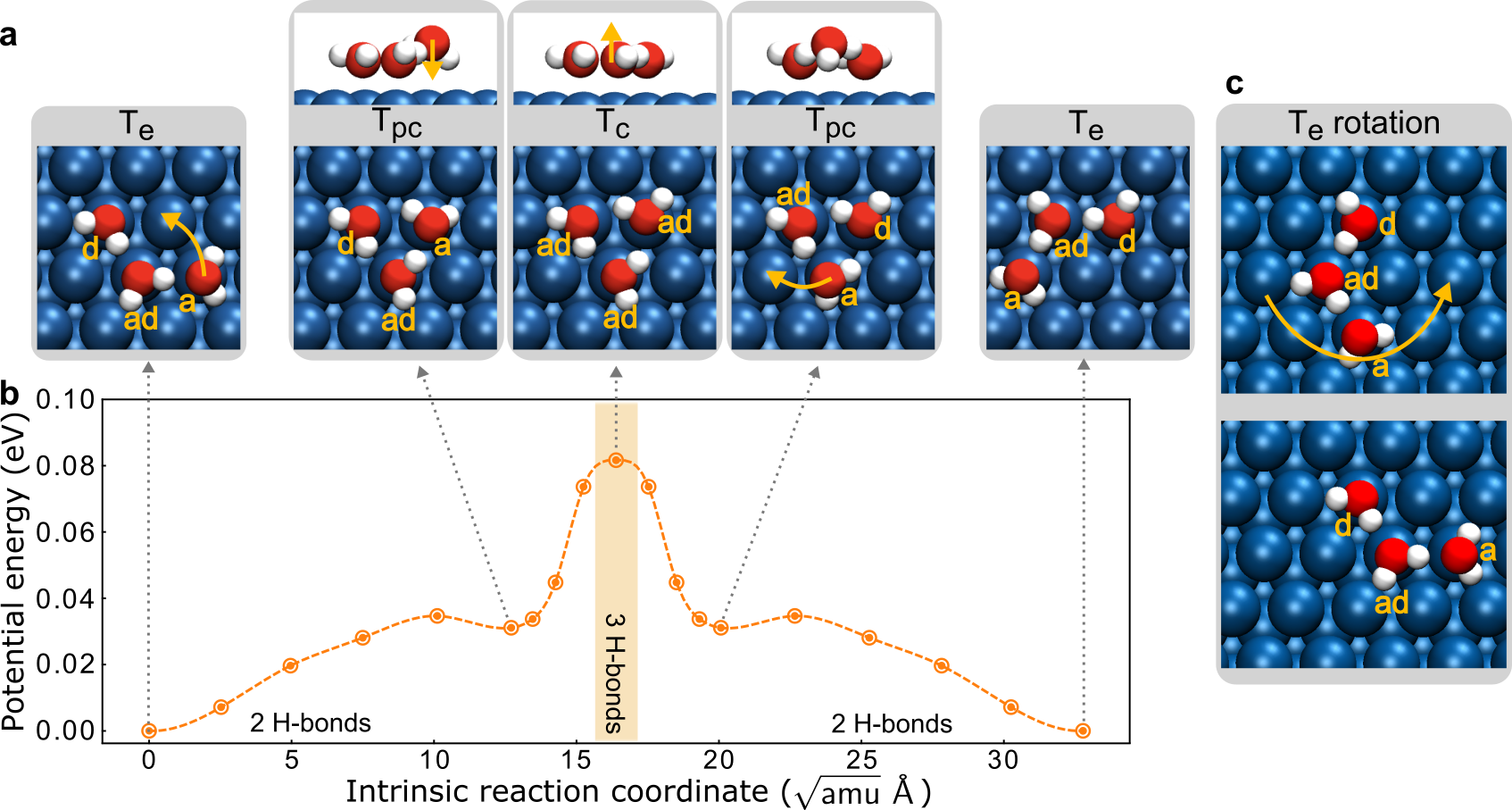}
\caption{A mechanism for facile water trimer motion.
(a) Top and side view of the inchworm mechanism.
(b) Minimum energy pathway characterized using the intrinsic reaction coordinate \cite{IRC1} for the inchworm diffusion mechanism of the water trimer.
(c) T$_\text{e}$ rotation process that completes the inchworm diffusion process.
\label{image_theory}} 
\end{figure}

%
%
However, analysis of the various trimer structures and their interconversion mechanisms offers important insights. 
The transition from the elongated to the pseudo-cyclic structure occurs simply through a rotational motion in which the $a$ molecule moves across a bridge site (Fig.\ \ref{image_theory}(a)). 
The H-bond between the $ad$ and $a$ molecules is retained during this process and as a consequence the 
$ad$ molecule rotates by ca. 60$^\circ$, while the position of the $d$ molecule is hardly affected.
Following this the pseudo-cyclic structure can further convert into a C$_3$-symmetric cyclic trimer (Fig.~\ref{image_theory}(a), middle).
If from this symmetric position the roles of the original 
$a$, $ad$, and $d$ molecules exchange (e.g.\ $d$ becomes $ad$, $ad$ becomes $a$, and $a$ becomes $d$) then when the trimer relaxes back to an elongated structure it will have moved by one surface atom.
Subsequently the $ad$ and $d$ waters in the trimer can rotate about the $d$ water thus completing the full diffusion process (Fig.~\ref{image_theory}(c)).
Thus the overall process involves an elongated trimer, converting to a cyclic trimer via the pseudo-cyclic trimer, which converts back to a different pseudo-cylcic trimer, which then opens to a different elongated trimer (Fig.\ref{image_theory}(a)).
The overall barrier for this process is approx. 80 meV (Fig.\ref{image_theory}(a) and Table I); a barrier that is lower than the monomer, dimer, or any of the other trimer diffusion barriers identified when all barriers are computed with a consistent set of computational settings \cite{dimersurf}.
Because the overall process involves contraction and expansion in a manner similar to inchworm propulsion, we refer to the mechanism as ``inchworm"-like diffusion.


It remains to be seen whether facile water timer diffusion can occur on transition metal surfaces in general or it is unique to Cu. 
We have shown that the barrier of the inchworm trimer diffusion mechanism mainly depends on the energy of the cyclic trimer T$_\text{c}$ relative to T$_\text{e}$. 
Therefore, as a first step towards understanding trimer diffusion on other metal surfaces, we computed the energies of T$_\text{c}$ and T$_\text{e}$ on the (111) surfaces of Ag, Pd, Pt, and Rh. 
The results show that on Pd, Pt, and Rh, the energy differences between T$_\text{c}$ and T$_\text{e}$ are 75 meV, 115 meV, and 84 meV respectively, which is smaller than the monomer and dimer diffusion barriers computed on these surfaces (with the same functional) in our previous work. \cite{dimersurf}
This suggests that the trimer could be more mobile than the dimer and the monomer on these metal surfaces as well. 
On Ag, the most stable conformer is T$_\text{pc}$, indicating that water trimers on this surface are different.

Before concluding, we briefly discuss the quantitative agreement between experiment and simulation in terms of energetics; an issue of broad interest to diffusion at surfaces. 
The inchworm mechanism rationalizes the experimental observations and the predicted DFT barrier of $ca.$ 80 meV is lower than the diffusion barriers for either the monomer or dimer. 
However, the barrier is higher than expected for a thermally activated process to occur at $<$10 K at the rates observed experimentally. 
To address this issue we first explored the sensitivity of the various diffusion barriers to the basic computational settings (unit cell size, surface relaxation) as well as the inclusion of ZPE effects. 
These results are shown in Table I where it can be seen that the ``standard'' and subsequent ``corrected'' barriers never differ by much (2-12 meV). 
In addition, tests with different DFT exchange-correlation functionals show that altering the functional does not lead to significant changes in the observed behavior (SI section S.VIII). 
Previous calculations have shown that nuclear tunneling can, for certain systems, play an important role in water cluster diffusion \cite{ranea04,dimersurf,GUO2017203}. 
However, a preliminary analysis shows that despite evidence of significant tunneling (including oxygen tunneling) in this process, these effects alone are insufficient to explain the low temperature mobility of the water trimers (SI section S.IX). 
Another effect that seems more likely to 
be relevant is the influence of the electric field from the STM tip. 
Tip effects are known to have an influence on adsorbate diffusion in general, including water diffusion \cite{PhysRevLett.101.136102,PhysRevMaterials.5.065001,KUMAGAI2015239}. 
In fact, recently it has been shown for water monomer diffusion on metals that applying a constant chemical potential for the electrons yields lower barriers and better quantitative agreement with experiments \cite{PhysRevMaterials.5.065001} and large bias induced water molecule reorientation can be visualized by STM \cite{mehlhorn14}.
Indeed, through careful examination of the STM time series obtained for the water trimers we do see some evidence that the electric field from the tip could play a role. 
Simulations (reported in SI section S.X) also reveal that an electric field can alter the barrier for the inchworm process; albeit to a minimal extent small ($<$ 10 meV). 
However, going beyond this and quantifying the influence of the tip with STM through e.g. temperature dependent measurements of the trimer diffusion process is highly challenging because the temperature difference between no observable motion and facile rotation plus diffusion is a mere 6 K.
In addition, it is not clear how the contributions of the individual processes could be deconvoluted experimentally. 
However, independent of a possible influence of the electric field on the extent of trimer motion, the direct comparison to the simultaneously imaged monomers and dimers is qualitatively consistent with the smaller trimer barrier deduced theoretically. 
Overall more work is required to fully resolve this issue and obtaining quantitative agreement for diffusion at surfaces remains a general challenge for future research.

\section{Conclusion}

In conclusion, we have demonstrated that water trimers show a much richer and faster kinetics than water monomers and other water clusters on Cu(111).
We have revealed that the interconversion between different configurations aided by a facile rotation leads to a center-of-mass motion 
far below the 
temperature at which monomers and dimers are observed to diffuse.
The overall interaction strength of the trimer with the surface is greater than that of either the monomer or dimer.
Thus we have found another example of how the delicate balance of hydrogen bonding and water-surface interactions can enable more strongly bound adsorbates to diffuse more rapidly than weakly bound adsorbates \cite{dimersurf}. 
In addition, in contrast to the water dimer, the motion of trimers occurs without the need for hydrogen bond exchange. 

Water cluster motion at the temperatures considered here demonstrates that the growth of extended ice structures may be easier than previously expected. 
Moreover, kinetic models for ice growth on surfaces should take growth through trimer and other water cluster building blocks into consideration. 
Indeed since the proportion of trimers on the surface will vary with temperature and depend on the mobility of monomers and dimers from which they form, fascinating growth kinetics is to be expected. 
Beyond ice, it would be interesting to explore if the exceptionally facile interconversion between different hydrogen bonded structures seen here on Cu(111) might go some way to explaining rapid water flow under nano-confinement \cite{siria_new_2017}.   
Finally, it is worth considering if the large angular flexibility of hydrogen bonded systems in general could be exploited to facilitate rapid motion of other types of hydrogen bonded clusters across the surfaces of materials.

\section{ASSOCIATED CONTENT}
The Supporting Information is available free of charge at https://pubs.acs.org/
\begin{itemize}
    \item STM experiment footage.
    \item Additional experimental and computational results.
\end{itemize}

\section{Acknowledgments}

This work was supported by the Research Training group 'Confinement-controlled Chemistry', which is funded by the Deutsche Forschungsgemeinschaft (DFG, German Research Foundation) - GRK2376 / 331085229 and by the DFG under Germany's Excellence Strategy - EXC-2033 - Projektnummer 390677874 RESOLV.
We are grateful to the Materials Chemistry Consortium (funded by EPSRC (EP/L000202)) and to the UK Materials and Molecular Modelling Hub (funded by EPSRC (EP/P020194/1 and EP/T022213/1)) for computational resources.  
\section{Author contributions}

K.M. and A.M. conceived and supervised the project. K.M., M.adH., and C.Z. performed the experiments and K.M. analyzed the experimental data. W.F. carried out the simulations and W.F. and A.M. analyzed the simulation data. A.M., K.M., and W.F. co-wrote the manuscript. All authors discussed the results and commented on the manuscript.

\bibliography{ref}

\includepdf[pages=-,pagecommand={},width=\textwidth]{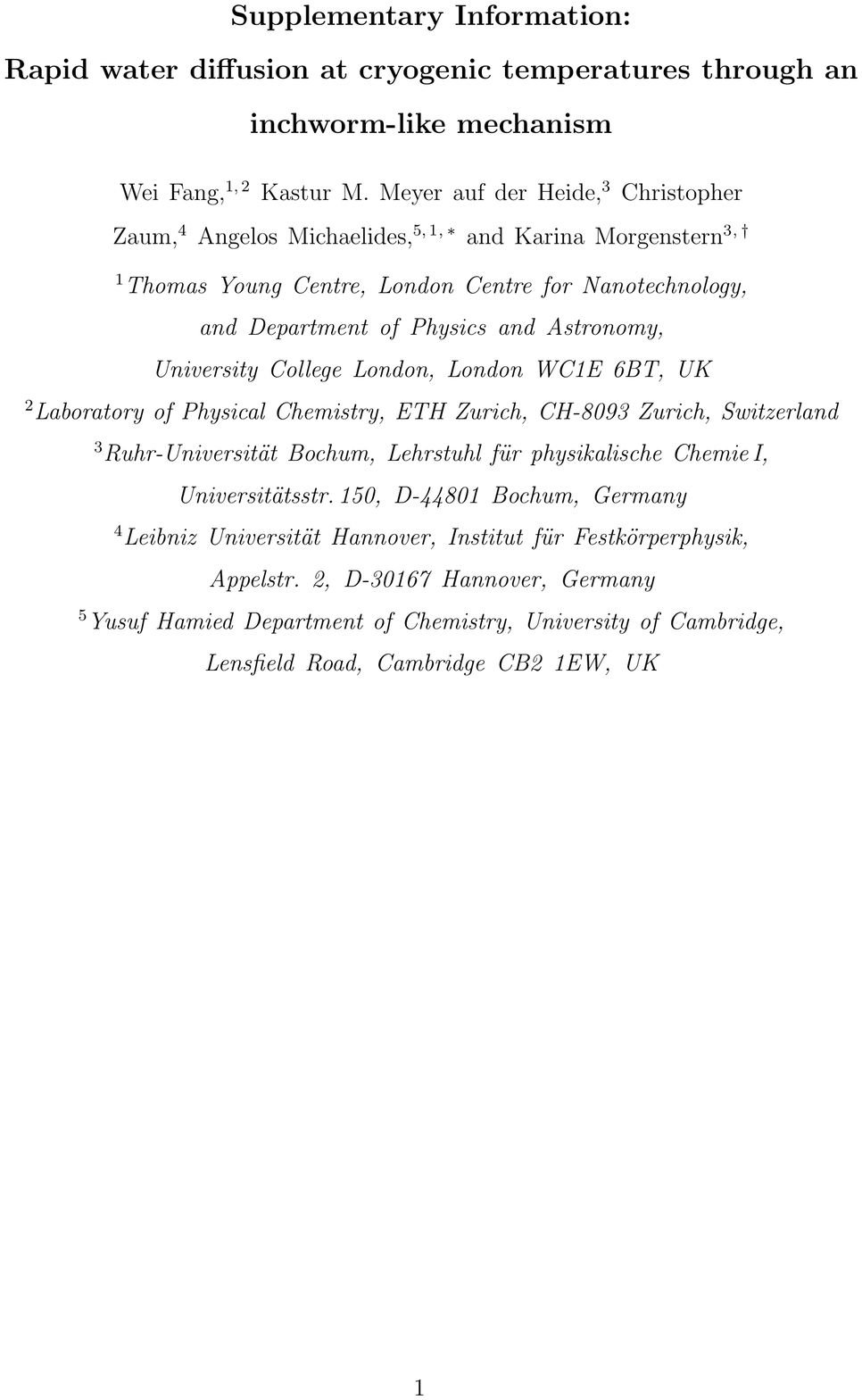}

\end{document}